\documentclass[12pt]{iopart}

\usepackage{iopams}
\usepackage{graphicx}

\expandafter\let\csname equation*\endcsname\relax
\expandafter\let\csname endequation*\endcsname\relax
\usepackage{amsmath}

\begin{document}

\title[Two-oscillator model of trapped-modes interaction]
{Two-oscillator model of trapped-modes interaction in a nonlinear
bilayer fish-scale metamaterial}

\author{Vladimir R Tuz$^{1,2}$, Bogdan A Kochetov$^{1}$, Lyudmila A Kochetova$^{3}$, Pavel L Mladyonov$^{1}$, and Sergey L Prosvirnin$^{1,2}$}

\address{$^1$ Institute of Radio Astronomy of National Academy of Sciences of Ukraine,
4, Krasnoznamennaya st., Kharkiv 61002, Ukraine}
\address{$^2$ School of Radio Physics, V.N.~Karazin Kharkiv National University, 4,
Svobody sq., Kharkiv 61022, Ukraine}
\address{$^3$ O.Ya.~Usikov Institute for Radiophysics and Electronics
of the National Academy of Sciences of Ukraine, 12, Proskury st.,
Kharkiv 61085, Ukraine} \ead{tvr@rian.kharkov.ua, bkochetov@bk.ru,
lakochetova@bk.ru, mladyon@rian.kharkov.ua, and
prosvirn@rian.kharkov.ua}

\begin{abstract}
We discuss the similarity between the nature of resonant
oscillations in two nonlinear systems, namely, a chain of coupled
Duffing oscillators and a bilayer fish-scale metamaterial. In such
systems two different resonant states arise which differ in their
spectral lines. The spectral line of the first resonant state has a
Lorentzian form, while the second one has a Fano form. This
difference leads to a specific nonlinear response of the systems
which manifests itself in appearance of closed loops in spectral
lines and bending and overlapping of resonant curves. Conditions of
achieving bistability and multistability are found out.
\end{abstract}

\noindent{\it Keywords\/}: coupled oscillators, Duffing oscillator,
metamaterial, trapped mode, bistability. \pacs{05.45.-a, 05.45.Xt,
42.25.Bs, 42.65.Pc, 78.67.Pt} \submitto{\PS} \maketitle

\section{Introduction} \label{sec:introduction}

The resonant phenomena are inherent to all types of vibrations or
waves and a number of resonant states in mechanic, acoustic,
electromagnetic and quantum systems are well known. Importantly,
there are some standard approaches for describing such different
resonant phenomena in various branches of physics which are
primarily developed within the oscillation theory framework. Within
this theory, despite the differences in the nature of the resonant
phenomena, they are described from a unified standpoint with similar
or even the same equations and techniques.

A resonance is thought to be an enhancement of the response of a
system to an external excitation at a particular frequency. It is
referred to the resonant frequency or natural frequency of the
system. From the oscillation theory standpoint a resonance is
introduced by the means of a harmonic oscillator being under an
action of a periodic driving force. When the frequency of the
driving force is close to the eigenfrequency of the oscillator, the
amplitude of oscillations grows toward its maximal value. Besides
that, many physical systems may also exhibit the opposite phenomenon
when their response is suppressed under certain resonant conditions
which sometimes is named with the term \textit{antiresonance}
\cite{miroshnichenko_RevModPhys_2010}. The simplest example can be
illustrated using two coupled harmonic oscillators, where one of
them is driven by a periodic force. Remarkably such a system of two
coupled harmonic oscillators simultaneously supporting
resonant-antiresonant states is considered as an intuitive and
popular model to describe features of many resonant phenomena,
including electromagnetically induced transparency (EIT)
\cite{garrido_AmJPhys_2002, harden_EurJPhys_2011}, stimulated
resonant Raman effect \cite{hemmer_JOSAB_1988}, level repulsion
\cite{frank_AmJPhys_1994}, conditions for adiabatic and diabatic
transitions \cite{shore_AmJPhys_2009, novotny_AmJPhys_2010}, quantum
coherence-decoherence \cite{magalhaes_PhysScr_2006}, etc.

Thereby in the system of two coupled oscillators, in general, there
are two resonances located close to the certain eigenfrequencies of
each oscillator \cite{joe_PhysScr_2006}. One of the resonances of
the forced oscillator demonstrates the standard amplitude growing
near its eigenfrequency and it has a \textit{symmetric} spectral
line, described by Lorentzian function. At the same time the other
resonance demonstrates an unusual sharp peak in the amplitude and it
has an \textit{asymmetric} spectral line, known as Fano profile. At
the antiresonant state there is a total suppression of the amplitude
of the forced oscillator at the eigenfrequency of the second
oscillator which is one of its basic properties, originated from the
resonant destructive interference, that distinguishes the Fano
resonance among the other ones \cite{joe_PhysScr_2006}.

The resonant-antiresonant states were originally studied in quantum
physics in relation to asymmetrically shaped ionization spectral
lines of atoms and molecules. But in the recent years they attract
appreciable attention in the field of plasmonic nanoparticles,
photonic crystals, and then electromagnetic metamaterials
\cite{miroshnichenko_RevModPhys_2010,
lukyanchuk_NatureMaterials_2010, khardikov_JOpt_2010,
khardikov_JOpt_2012, khanikaev_Nanoph_2013}. This interest is
stimulated by promising applications of resonances with asymmetric
spectral lines in sensors, lasing, switching, nonlinear and slow
light devices, due to the steep dispersion of their profile. Despite
the fact that the nature of such resonances in photonic devices is
quite complicated and it is explained by the interference effect
between a certain non-radiative mode and a continuum of radiative
electromagnetic waves, the simple two-oscillator model is still
widely used to reveal the main resonant features of the optical
systems \cite{gallinet_PhysRevB_2011, tassin_PhysRevLett_2012,
taubert_JOSAB_2013}.

Another important characteristic of structures supporting
resonant-antiresonant states is their possibility to provide an
enhanced energy storing. Remarkably, simultaneous presence of both
steep resonant feature and strong field localization brings a
possibility of realization an optimal bistable switching in
nonlinear systems. In particular, in optical systems, the main idea
of using the resonant-antiresonant states for all-optical switching
and bistability is to introduce an element with nonlinear
characteristic and achieve a stepwise nonlinearly-induced shift of
the resonant frequency \cite{soljacic_PhysRevE_2002,
cowan_PhysRevE_2003, maes_JOSAB_2005, wurtz_PhysRevLett_2006,
tuz_PhysRevB_2010, tuz_EurPhysJ_2011, tuz_JOpt_2012}. Thus, by
employing such nonlinear shift one can reach bistability in many
devices suggested on the plasmonic, photonic crystal and
metamaterial platforms.

From the  viewpoint of the oscillation theory, a study of
resonant-antiresonant states in nonlinear systems causes derivation
of the particular model of a chain of two \textit {nonlinearly}
coupled oscillators \cite{lifshitz_PhysRevB_2003}. In a mathematical
form such a system can be described by a set of two coupled Duffing
equations which is the basic model for illustrating synchronization
phenomenon and related effects \cite{afraimovich_PhysD_1997,
vincent_PhysScr_2008, kuznetsov_PhysD_2009}. It is known that,
despite the apparent simplicity of the Duffing equations, there are
no easy ways to find an exact analytical solution for the
corresponding system of nonlinear equations. In this regard, for its
solution some asymptotic approaches are traditionally used. Howbeit
obtained solution can comprise a number of peculiarities and stand
out by the presence of hysteresis, several stable cycles, complex
dynamics and chaotic regimes.

The complete study of the system of nonlinear equations supposes
involving a concept of dynamical systems \cite{woafo_PhysScr_1998}.
However, in this paper we intend to restrict ourself only to
extension the results of \cite{joe_PhysScr_2006} by adding a weak
nonlinearity to the system of two coupled oscillators and solving it
with the slowly varying amplitude approximation in the frequency
domain. The main purpose of the paper is to reveal general changes
in the spectral line of the resonant-antiresonant states when such
weak nonlinearity is introduced into the system.

Then on the basis of the results obtained from the nonlinear
two-oscillator model we propose an example of a nonlinear
metamaterial configuration which operating regimes qualitatively
resemble characteristics of the mentioned oscillating system. As
such a structure we consider a special class of metamaterials which
involves  planar metasurfaces supporting so-called ``trapped-modes".
The trapped-mode is a specific resonant state that appears in the
metamaterials made of subwavelength metallic or dielectric particles
(inclusions) with a certain asymmetric form
\cite{khardikov_JOpt_2010, khardikov_JOpt_2012,
prosvirnin_NATO_2003, fedotov_PhysRevLett_2007}. The trapped-modes
are the result of \textit {antiphase} oscillations of fields on the
particles parts (arcs) and are excited by an external
electromagnetic field. In literature such trapped-mode metamaterials
sometimes are also referred to EIT-like metamaterials
\cite{papasimakis_PhysRevLett_2008}. It is due to the fact that
their response is a direct classical analog of EIT because the weak
coupling of the antiphased local fields to free space is reminiscent
of the weak probability for photon absorption in EIT observed in
atomic system. In this paper we consider a particular configuration
of such metamaterial, namely, a nonlinear bilayer fish-scale
structure \cite{fedotov_PhysRevE_2005}.

\section{Two-oscillator model: Set of coupled Duffing equations}
\label{sec:oscillators}
Our objective here is to study the main spectral features of a chain
of two coupled nonlinear oscillators. For this reason we consider
the two harmonic oscillator system which models classically the Fano
resonance \cite{joe_PhysScr_2006}. If it is supplemented by the
cubic nonlinear terms, we arrive to the set of two coupled Duffing
equations related to the coordinates $x_1$ and $x_2$ in the form
\begin{equation} \label{System}
\begin{array}{l}
{\dfrac{d^2x_1}{dt^2}}+2\delta_1\dfrac{dx_1}{dt}+\omega^2_1x_1+
\omega^2_1\beta_1x^3_1-c_1x_2=P_0\cos(\omega t), \\
{}\\
{\dfrac{d^2x_2}{dt^2}}+2\delta_2\dfrac{dx_2}{dt}+\omega^2_2x_2+
\omega^2_2\beta_2x^3_2-c_2x_1=0, \\
\end{array}
\end{equation}
where $\delta_1$ and $\delta_2$ are the damping coefficients,
$\omega_1$ and $\omega_2$ are the natural frequencies, $\beta_1$ and
$\beta_2$ are the nonlinear coefficients, and the coupling between
the oscillators is characterized by the coefficients $c_1$ and
$c_2$. The first oscillator ($x_1$) is submitted to the action of an
external harmonic force with amplitude $P_0$ and frequency $\omega$.

As it is generally characteristic of actual nonlinear systems, weak
nonlinearity ($\beta_1\ll 1$, $\beta_2\ll 1$), weak coupling
($c_1\ll 1$, $c_2\ll 1$) and low damping ($\delta_1\ll 1$,
$\delta_2\ll 1$) can be supposed. Additionally we assume that the
driving harmonic force has a small amplitude ($P_0\ll 1$), and the
resonant frequencies and the frequency of the driving force are
closely spaced ($\omega_1\sim\omega_2\sim\omega$). Under such
quasi-linear conditions the method of slowly varying amplitude
\cite{mickens_1981} can be applied to solve the set of nonlinear
equations (\ref{System}). In the framework of this method the
solution of the system (\ref{System}) is sought in the form
$x_1(t)=A\cos(\omega t+\theta)$ and $x_2(t)=B\cos(\omega t+\phi)$,
where $A$ and $B$ are the slowly varying amplitudes, and $\theta$
and $\phi$ are the slowly varying phases.

Making the standard change of variables \cite{mickens_1981} and
subsequent averaging we arrive to the following system of reduced
equations
\begin{equation} \label{RedEq}
\begin{array}{l}
{\dfrac{dA}{d\tau}}=-A-k_1B\sin{(\theta-\phi)}-P\sin{\theta}, \\
{}\\
{A\dfrac{d\theta}{d\tau}}=-\Omega A+\gamma_1A^3-k_1B\cos{(\theta-\phi)}-P\cos{\theta}, \\
{} \\
{\dfrac{dB}{d\tau}}=-\delta B-k_2A\sin{(\phi-\theta)}, \\
{} \\
{B\dfrac{d\phi}{d\tau}}=-(\Omega-\eta)B+\gamma_2B^3-k_2A\cos{(\phi-\theta)}, \\
\end{array}
\end{equation}
where $\tau=\delta_1 t$ is the ``slow'' time,
$\gamma_1=3\omega\beta_1/8\delta_1$ and
$\gamma_2=3\omega\beta_2/8\delta_1$ are the normalized nonlinear
coefficients, $k_1=c_1/2\omega\delta_1$ and
$k_2=c_2/2\omega\delta_1$ are the normalized coupling coefficients,
$\delta=\delta_2/\delta_1$ is the relative damping,
$\Omega=(\omega^2-\omega_1^2)/2\omega\delta_1$ is the frequency
mismatch, $\eta=(\omega_2^2-\omega_1^2)/2\omega\delta_1$ is the
frequency difference and $P=P_0/2\omega\delta_1$ is the normalized
amplitude of the driving force. The set of equations (\ref{RedEq})
has a steady-state solution as $\tau\to\infty$. Practically, to find
the steady-state solution we use the conditions of constant
amplitudes and phases, i.e. we set
$dA/d\tau=dB/d\tau=d\theta/d\tau=d\phi/d\tau=0$. Thus, the
steady-state solution of the system (\ref{RedEq}) satisfies the
following system of algebraic equations
\begin{equation} \label{StatEq}
\begin{array}{l}
{A^2=\dfrac{B^2}{k_2^2}\left[\delta^2+(\Omega-\eta-\gamma_2B^2)^2\right],} \\
{}\\
\left[A^2+\dfrac{k_1}{k_2}\delta B^2\right]^2+
\left[\Omega A^2-\gamma_1A^4-\dfrac{k_1}{k_2}B^2
\left(\Omega-\eta-\gamma_2B^2\right)\right]^2=A^2P^2. \\
\end{array}
\end{equation}
It defines the steady-state amplitudes of small nonlinear
oscillations existing in the set of two coupled Duffing oscillators
(\ref{System}).

As it is well known \cite{joe_PhysScr_2006}, if the nonlinearity in
the system (\ref{System}) is absent ($\beta_1=\beta_2=0$) each of
the two linear oscillators has two resonant states at some
particular frequencies. In such a linear system the amplitudes can
be calculated exactly using the complex amplitude method. Assuming
notations used here they have the following form
\begin{equation} \label{LinEq}
A=\left|\dfrac{\left(-\Omega+\eta+\mathrm{i}\delta\right)P}
{\left(\Omega-\mathrm{i}\right)\left(\Omega-\eta-\mathrm{i}
\delta\right)-k_1k_2}\right|,
\qquad
B=\left|\dfrac{k_2P}{\left(\Omega-\mathrm{i}\right)
\left(\Omega-\eta-\mathrm{i}\delta\right)-k_1k_2}\right|.
\end{equation}
It should be noted that in order to simplify derivation of the
amplitudes (\ref{LinEq}) we use a complex harmonic exponent instead
of a real cosine function in the right hand side of the first
equation in (\ref{System}).

\begin{figure}[htb]
\centerline{\includegraphics[width=9.0cm]{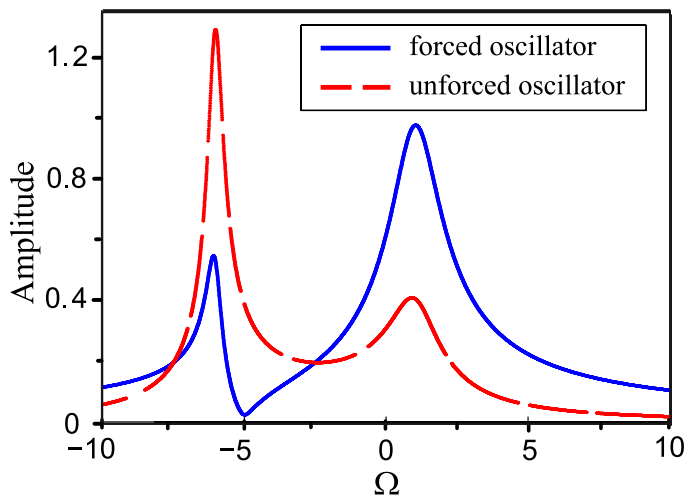}}
\caption{Resonant features of amplitudes of two coupled oscillators
in the linear system; $P=1$, $\delta=0.15$, $\eta=-5$,
$k_1=k_2=2.5$, $\gamma_1=\gamma_2=0$.} \label{fig:fig1}
\end{figure}

Typical dependences of the amplitudes (\ref{LinEq}) on the frequency
$\Omega$ are presented in Fig.~\ref{fig:fig1}. The spectral line of
the second linear oscillator (which is unforced) has two resonances
with symmetrical Lorentzian shape positioned nearly the
corresponding eigenfrequencies (see the red dash line in
Fig.~\ref{fig:fig1}). At once the spectral line of the first
oscillator (which is forced) has two resonances with both
symmetrical Lorentzian and asymmetrical Fano shape (see the blue
solid line in Fig.~\ref{fig:fig1}). The Fano resonance is a result
of the composition (interference) of two oscillations from the
driving force and the second coupled oscillator. If the phase of the
oscillator changes monotonously when the driving frequency passes
through the resonance the Lorentzian resonance is observed. On the
contrary if the phase dependence has a gap then the Fano resonance
appears \cite{joe_PhysScr_2006}. In particular for the considered
system the antiresonant state takes place at the frequency $\Omega =
\eta = -5$.

\begin{figure}[htb]
\centerline{\includegraphics[width=9.0cm]{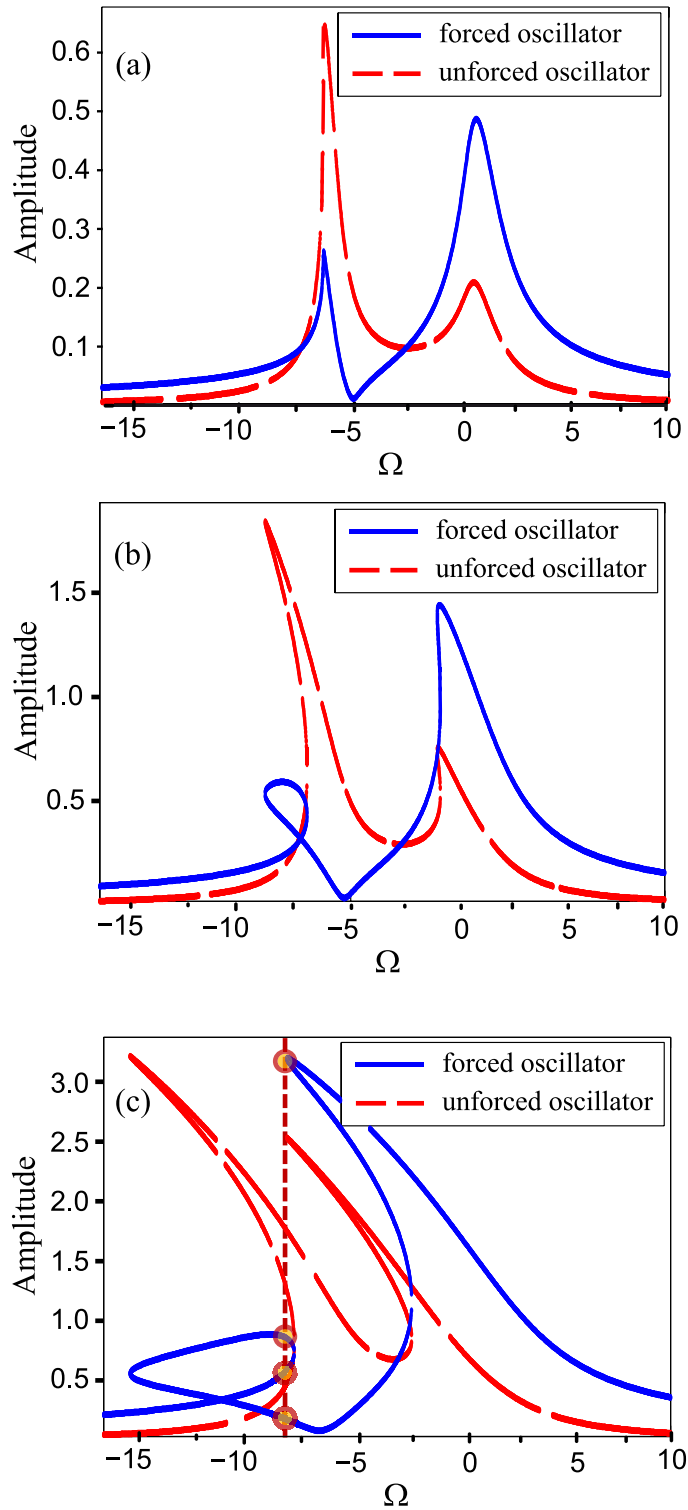}}
\caption{Amplitude-frequency responses of coupled nonlinear
oscillators for different amplitudes of the driven force $P$;
$\delta=0.15$, $\eta=-5$, $k_1=k_2=2.5$, $\gamma_1=\gamma_2=-1$; (a)
$P=0.5$; (b) $P=1.5$; (c) $P=3.5$.} \label{fig:fig2}
\end{figure}

The shape of the linear resonances does not depend on the amplitude
of the driving force. In other words, changing the amplitude $P$ in
equations (\ref{LinEq}) only leads to scaling along the ordinate
axis in Fig.~\ref{fig:fig1}. But it is not the same in the case of
nonlinear system ($\beta_1\ne 0$, $\beta_2 \ne 0$). Indeed, one can
see the complicated dependences of the steady-state amplitudes on
the driving force in equations (\ref{StatEq}). Peculiarity of the
nonlinear resonance consists in the presence of significant
dependence of the resonant shapes on the amplitude of the driving
force. Such frequency dependences of the steady-state amplitudes
(\ref{StatEq}) on the driving force $P$ are represented in
Fig.~\ref{fig:fig2}. These curves illustrate the transformations of
the coupled Lorentzian and Fano resonances in the system
(\ref{System}) having weak nonlinearity. For the small amplitudes of
the driving force ($P=0.5$) there is no significant difference
between the nonlinear resonances features and linear ones. Further
increasing the amplitude of the driving force leads to some
deformation of the resonant curves. The peaks of Lorentzian
resonances become frequency shifted and their shape gets bended. The
form of the spectrum line of the left resonant state transforms into
a closed loop while the frequency of the antiresonant state acquires
a shift [Fig.~\ref{fig:fig2}(b),(c)].

It is known that for each nonlinear resonance there is some critical
amplitude of the driving force. If the amplitude of the driving
force is greater than the critical one then points with a vertical
tangent line appear on the resonant curve. It should be noted that
the left and the right resonances of the same resonant curve have
different critical amplitudes. But the corresponding resonances of
the different curves have the same critical amplitudes. When the
amplitude of the driving force is high enough the frequency bands
appear where the steady-state amplitudes have an ambiguous
dependence (hysteresis) on the frequency of the driving force. Some
parts of the resonant curves within these bands become unstable
sets. Therefore, continuous variation of the driving force frequency
leads to jumping of the steady-state amplitudes on the boundaries of
the unstable regions which results in appearance of bistability.

For the certain parameters of the system (\ref{System}) overlapping
of the two different resonances can be observed. Evidently, in this
case, the spectral curves acquire more than two stable states (which
are marked in Fig.~\ref{fig:fig2}(c) with circles), i.e., the effect
of multistability arises in the system.

\section{Electromagnetic analog: Nonlinear bilayer fish-scale metamaterial}
\label{sec:metamaterial}
Subsequently, in this section, we confirm
the predictions of our nonlinear two-oscillator model in an optical
system. As such a system we consider a particular configuration of a
metamaterial in which resonant-antiresonant states can be excited
effectively using trapped-modes. It consists of equidistant arrays
of continuous meander metallic strips placed on both sides of a thin
dielectric substrate (bilayer fish-scale structures
\cite{fedotov_PhysRevE_2005}). In such a fish-scale structure the
trapped mode resonances can be exited if the incident field is
polarized along the strips and when the form of these strips is
slightly different from the straight line. Furthermore, in the
bilayer structure, besides the trapped-mode resonance excited within
each grating, another trapped mode resonance can appear due to a
specific interaction of the antiphase current oscillations between
two adjacent gratings. Thereby our structure supports two distinct
resonant states which corresponds to the characteristic of the
two-oscillator model.

\begin{figure}[htb]
\centerline{\includegraphics[width=10.0cm]{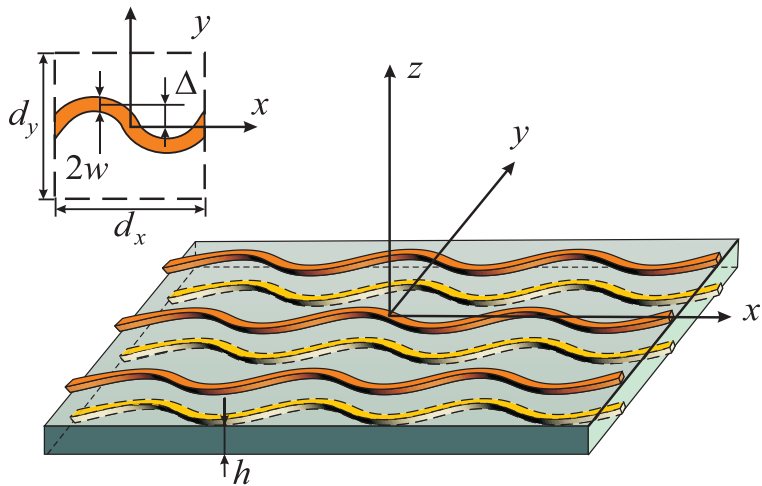}}
\caption{Fragment of a planar bilayer fish-scale metamaterial and
its unit cell.} \label{fig:sketch}
\end{figure}

A sketch of the studied structure is presented in
Fig.~\ref{fig:sketch}. The structure consists of two gratings of
planar perfectly conducting infinite wavy-line strips placed on both
sides of a dielectric slab with thickness $h$ and permittivity
$\varepsilon$. The elementary translation cell of the structure
under study is a square with sides $d=d_x=d_y$. The full length of
the strip within the elementary translation cell is $S$. Suppose
that the thickness $h$ and size $d$ are less then the wavelength
$\lambda$ of the incident electromagnetic radiation ($h\ll\lambda$,
$d<\lambda$). The width of the metal strips and their deviation from
the straight line are $2w$ and $\Delta$, respectively.

Assume that the normally incident field is a plane monochromatic
wave polarized parallel to the strips ($x$-polarization), and the
amplitude of the primary field is $A_0$.

In the frequency domain we use the method of moments to solve the
problem of electromagnetic wave scattering by the bilayer fish-scale
metamaterial \cite{prosvirnin_JElectromagWavesAppl_2002,
mladyonov_PhysicsAstronomy_2010}. It involves solving the integral
equation related to the surface currents which are induced in the
metal pattern by the field of the incident wave. In the framework of
the method of moments, the metal pattern is treated as a perfect
conductor, while the substrate is assumed to be a lossy dielectric.
In the bilayer configuration the method of solution rigorously takes
into account an electromagnetic coupling between two adjacent
gratings via evanescent partial spatial waves. The metamaterial
response can be expressed through the induced currents $J_1$ and
$J_2$ which flow along the strips of the corresponding grating and
the reflection $R$ and transmission $T$ coefficients  as functions
of normalized frequency ($\ae=d/l$), permittivity ($\varepsilon$)
and other parameters of the structure.

\begin{figure}[htb]
\centerline{\includegraphics[width=16.0cm]{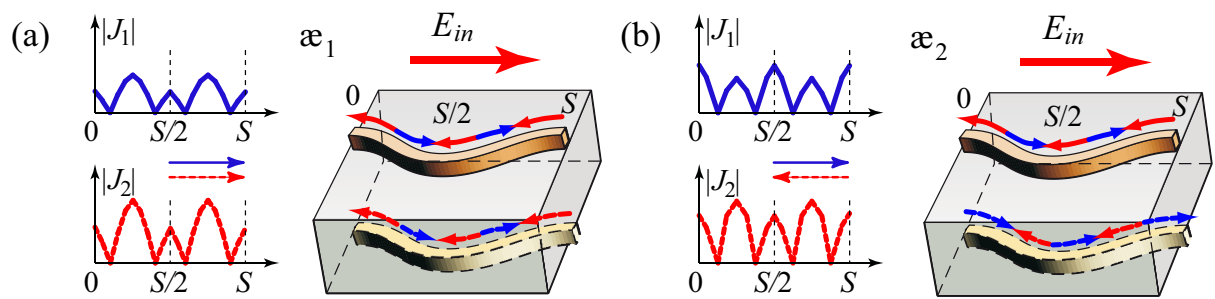}} \caption{The
surface current distribution along the strips placed on the upper
and bottom sides of the substrate in the bilayer fish-scale
metamaterial.} \label{fig:currents}
\end{figure}

Remarkably, due to the bilayer configuration of the structure under
study, there are two possible current distributions which cause the
trapped mode resonances. The first distribution is the antiphase
current oscillations in arcs of each grating. The currents flow in
the same manner on both gratings and the resonance exists due to the
curvilinear form of the strips. This resonance is inherent to both
single-layer and bilayer structure's configurations
\cite{prosvirnin_JElectromagWavesAppl_2002,
mladyonov_PhysicsAstronomy_2010}. The resonant frequency is labeled
in Fig.~\ref{fig:currents}(a) by the letter $\ae_1$, and it
corresponds to the first resonant frequency of our two-oscillator
model. The second distribution is the antiphase current oscillations
excited between two adjacent gratings. Obviously this resonance can
be excited only in the bilayer structure's configuration. The
resonant frequency is labeled in Fig.~\ref{fig:currents}(b) by the
letter $\ae_2$,  and hence it corresponds to the second resonant
frequency of the two-oscillator model.

\begin{figure}[htb]
\centerline{\includegraphics[width=9.0cm]{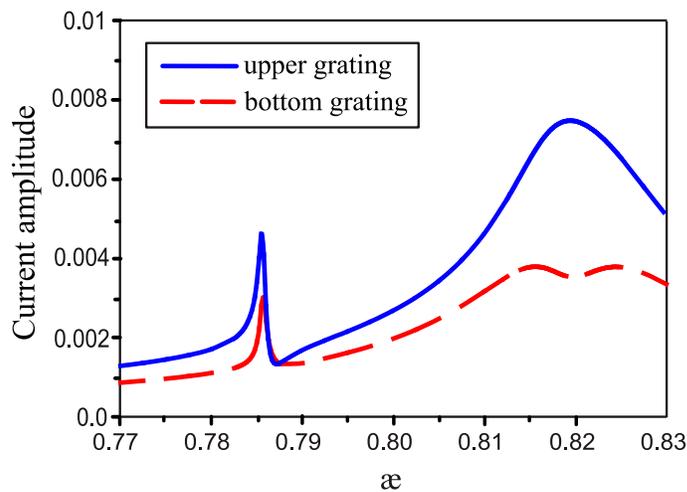}} \caption{The
frequency dependences $(\ae=d/\lambda)$ of current amplitudes
induced on the upper and bottom gratings; $\varepsilon = 3$,
$2w/d=0.05$, $h/d=0.2$, $\Delta/d=0.25$.} \label{fig:linspectra}
\end{figure}

Thereby the current oscillations on the upper and bottom gratings
are characterized with two resonant states which amplitudes are
presented in Fig.~\ref{fig:linspectra}. In contrast to the spectral
line of the current amplitude related to the bottom grating, the
corresponding spectral line related to the upper grating has a
specific asymmetric form with antiresonant state that is in full
compliance with predictions of the two-oscillator model. However a
small dip in the amplitude of the Lorentzian resonance related to
the bottom grating is explained by its incomplete screening with the
upper grating.

For a particular bilayer fish-scale structure, two resonant states
correspond to two peaks of reflectivity while the antiresonant state
corresponds to the maximum of transmissivity. At once these two
resonant states have different quality factors. The quality factor
of the first resonance depends on the form of strips and is
practically independent of the substrate permittivity $\varepsilon$.
Thus the less the form of strips is different from the straight
line, the greater is the quality factor of the first trapped-mode
resonance. On the other hand, the quality factor of the second
resonance crucially depends on both the distance between gratings
and permittivity of the substrate. Thus varying the distance between
gratings and substrate permittivity changes the trapped-mode
resonant conditions and this changing manifests itself in the
current amplitudes $J_1$ and $J_2$. We argue that, due to such
current distributions the field turns out to be localized between
the gratings, i.e. directly in the substrate, which can sufficiently
enhance the nonlinear effects if the substrate is made of a field
intensity dependent (nonlinear) material.

In this case, permittivity of the substrate $\varepsilon$ becomes to
be dependent on the intensity of the electromagnetic field inside it
($\varepsilon=\varepsilon_1+\varepsilon_2|E_{in}|^2$). In
\cite{tuz_PhysRevB_2010, tuz_EurPhysJ_2011, tuz_JOpt_2012,
tuz_JOSAB_2014} an approximate treatment was proposed to solve such
a nonlinear problem. It is obtained by introducing two
approximations. The first one postulates that the inner field
intensity is directly proportional to the square of the current
amplitude averaged over a metal pattern extent, $I_{in}\sim\Bar
J^2$, where $J = (J_1+J_2)/2$. The second approximation assumes
that, in view of the smallness of the elementary translation cell of
the array ($d < \lambda$), the nonlinear substrate remains to be a
homogeneous dielectric slab under an action of the intensive light.
These approximations allow us obtaining a nonlinear equation related
to the current amplitude averaged over metal pattern extent within
an elementary translation cell. The input field amplitude $A_0$ is a
parameter of this nonlinear equation. So, at a fixed frequency
$\ae$, the solution of this equation gives us the averaged current
amplitude $\bar J$ which depends on the amplitude of the incident
field $A_0$. On the basis of the current $\bar J(A_0)$ found by a
numerical solution of the nonlinear equation, the actual value of
permittivity $\varepsilon$ of the nonlinear substrate is determined
and the reflection $R$ and transmission $T$ coefficients can be
calculated as functions of the frequency $\ae$ and the amplitude of
the incident field $A_0$. For further details about the method of
solution the reader is referred to \cite{tuz_JOSAB_2014}.

\begin{figure}[htb]
\centerline{\includegraphics[width=9.0cm]{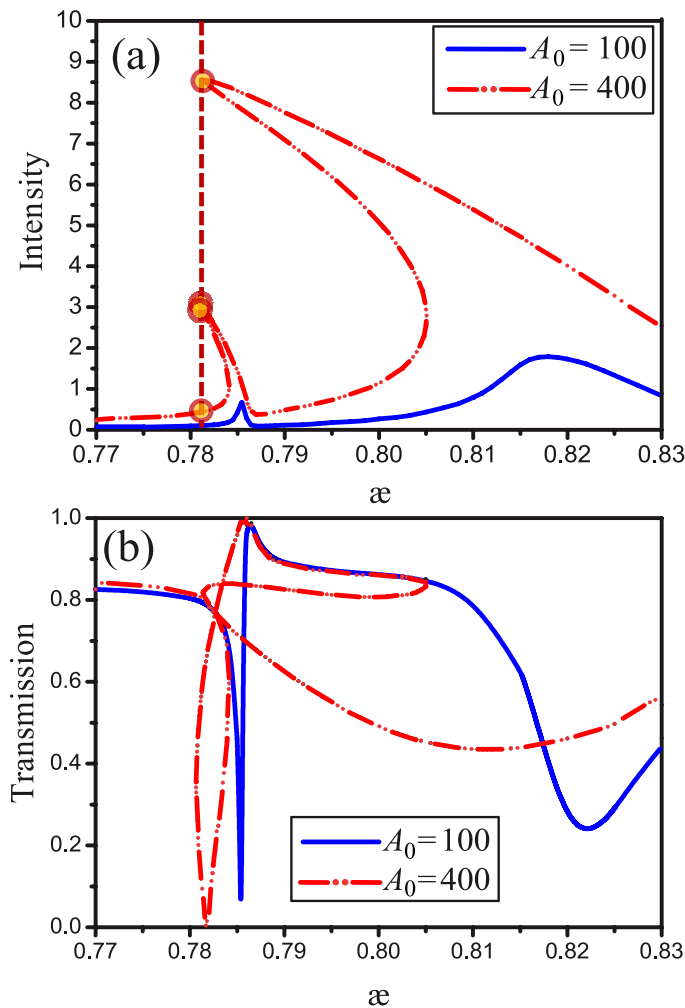}} \caption{The
frequency dependences $(\ae=d/\lambda)$ of (a) the inner field
intensity and (b) the transmission coefficient amplitude;
$\varepsilon_1=3$, $\varepsilon_2=0.005$~cm$^2$~kW$^{-1}$. Other
parameters are the same as in Fig.~\ref{fig:linspectra}.}
\label{fig:nonlinspectra}
\end{figure}

One can see that as the amplitude of the incident field rises, the
frequency dependences of the inner field intensity acquire a form of
bent resonances (Fig.~\ref{fig:nonlinspectra}) which completely
confirm the assumption of our nonlinear two-oscillator model. As
mentioned above, such form of lines is a result of the
nonlinearly-induced shift of the resonant frequency. In particular,
in the optical system, when the frequency of the incident wave is
tuned nearly the resonant frequency, the field localization produces
growing the inner light intensity which can alter the permittivity
enough to shift the resonant frequency \cite{tuz_JOSAB_2014}. When
this shift brings the excitation closer to the resonant condition,
even more field is localized in the system, which further enhances
the shift of resonance. This positive feedback leads to formation of
the hysteresis loop in the inner field intensity with respect to the
incident field amplitude, and, as a result, under a certain
amplitude of the incident field, the frequency dependences of the
inner field intensity take a form of bent resonances. Evidently, in
the certain frequency bands, the transmission coefficient acquires
two stable states where the effect of bistability takes place.

An important point is that in our system this bending is different
for distinctive resonances due to difference in their nature and,
respectively, in their current amplitudes \cite{tuz_JOSAB_2014}. It
results in a specific distortion of the curves of the transmission
coefficient amplitude nearly the trapped-mode resonant frequencies.
Thus, at the frequency $\ae_1\sim0.78$ the inner field produced by
antiphase current oscillations is confined in the area in the
vicinity of each grating and it weakly affects on the permittivity
of dielectric substrate. In this case the resonant line acquires a
transformation into a closed loop which is a dedicated
characteristic of sharp nonlinear Fano-shaped resonances and is
related to the characteristic of our nonlinear two-oscillator model.
The second resonance $\ae_2\sim0.82$ is smooth but the current
oscillations produce the strong field concentration between two
adjacent gratings directly inside the dielectric substrate. It leads
to a considerable distortion of the transmission coefficient
amplitude in a wide frequency range, and at a certain incident field
amplitude this resonance reaches the first one and tends to overlap
it [Fig.~\ref{fig:nonlinspectra}(b)]. Thus, in this case, the
transmission coefficient acquires more than two stable states, i.e.
the effect of multistability arises in full accordance to the
prediction of the two-oscillator  model.

\section{Conclusions}
\label{sec:conclusions}

In the present paper a direct analogy in oscillation characteristics
of two nonlinear systems is evolved. As such systems a chain of
coupled Duffing oscillators and an optical structure in the form of
bilayer fish-scale metamaterial  bearing trapped modes are
considered. It is shown that the spectral features of both systems
are distinguished by two resonant and single antiresonant states
which profiles acquire Lorentzian and Fano forms, correspondingly.

Certain peculiarities of nonlinear impact on spectral lines changing
with rising the amplitude of the driving force and the intensity of
the incident field are studied for the two-oscillator model and the
planar metamaterial, respectively. In the nonlinear regime,
resonance bending, closed loop formation, effects of bistability and
multistability in the spectra of both structures have been
demonstrated.

We argue that our nonlinear two-oscillator model can be used to
reveal physical nature of the resonant behavior of such a complicate
optical system and can help to identify conditions for the
appearance in it of chaotic oscillations, synchronization
phenomenon, and another related nonlinear effects.

\bigskip

The authors acknowledge the National Academy of Sciences of Ukraine
and the Ukrainian State Foundation for Basic Research for their
support with the project $\Phi$54.1/004.

\end{document}